\documentclass[12pt]{article}
\usepackage{fancyhdr}

\usepackage{mathrsfs}
\usepackage[T1]{fontenc}
\usepackage{mathpazo}
\usepackage{setspace}
\usepackage{amsfonts}
\usepackage{amssymb}
\usepackage{amsmath}
\usepackage{epsfig}
\usepackage{latexsym}
\usepackage{color}
\usepackage{graphicx}
\usepackage{nicefrac}
\usepackage[latin1]{inputenc}
\usepackage{slashed}
\usepackage{multirow}
\usepackage{comment}
\usepackage{soul}
\usepackage{hyperref}
\usepackage{cite}

\usepackage[titletoc]{appendix}

\def\hybrid{\topmargin -20pt    \oddsidemargin 0pt
        \headheight 0pt \headsep 0pt
        \textwidth 6.25in       
        \textheight 9.25in       
        \marginparwidth .875in
        \parskip 5pt plus 1pt   \jot = 1.5ex}

\hybrid

\def\baselinestretch{1.2}

\catcode`\@=11

\def\marginnote#1{}
\newcount\hour
\newcount\minute
\newtoks\amorpm
\hour=\time\divide\hour by60
\minute=\time{\multiply\hour by60 \global\advance\minute by-\hour}
\edef\standardtime{{\ifnum\hour<12 \global\amorpm={am}%
        \else\global\amorpm={pm}\advance\hour by-12 \fi
        \ifnum\hour=0 \hour=12 \fi
        \number\hour:\ifnum\minute<10 0\fi\number\minute\the\amorpm}}
\edef\militarytime{\number\hour:\ifnum\minute<10 0\fi\number\minute}

\def\draftlabel#1{{\@bsphack\if@filesw {\let\thepage\relax
   \xdef\@gtempa{\write\@auxout{\string
      \newlabel{#1}{{\@currentlabel}{\thepage}}}}}\@gtempa
   \if@nobreak \ifvmode\nobreak\fi\fi\fi\@esphack}
        \gdef\@eqnlabel{#1}}
\def\@eqnlabel{}
\def\@vacuum{}
\def\draftmarginnote#1{\marginpar{\raggedright\scriptsize\tt#1}}

\def\draft{\oddsidemargin -.5truein
        \def\@oddfoot{\sl preliminary draft \hfil
        \rm\thepage\hfil\sl\today\quad\militarytime}
        \let\@evenfoot\@oddfoot \overfullrule 3pt
        \let\label=\draftlabel
        \let\marginnote=\draftmarginnote
   \def\@eqnnum{(\theequation)\rlap{\kern\marginparsep\tt\@eqnlabel}%
\global\let\@eqnlabel\@vacuum}  }

\def\preprint{\twocolumn\sloppy\flushbottom\parindent 2em
        \leftmargini 2em\leftmarginv .5em\leftmarginvi .5em
        \oddsidemargin -.5in    \evensidemargin -.5in
        \columnsep .4in \footheight 0pt
        \textwidth 10.in        \topmargin  -.4in
        \headheight 12pt \topskip .4in
        \textheight 6.9in \footskip 0pt
        \def\@oddhead{\thepage\hfil\addtocounter{page}{1}\thepage}
        \let\@evenhead\@oddhead \def\@oddfoot{} \def\@evenfoot{} }

\def\numberbysection{\@addtoreset{equation}{section}
        \def\theequation{\thesection.\arabic{equation}}}

\def\underline#1{\relax\ifmmode\@@underline#1\else
        $\@@underline{\hbox{#1}}$\relax\fi}

\def\titlepage{\@restonecolfalse\if@twocolumn\@restonecoltrue\onecolumn
     \else \newpage \fi \thispagestyle{empty}\c@page\z@
        \def\thefootnote{\fnsymbol{footnote}} }

\def\endtitlepage{\if@restonecol\twocolumn \else \newpage \fi
        \def\thefootnote{\arabic{footnote}}
        \setcounter{footnote}{0}}  

\catcode`@=12
\relax

\def\figcap{\section*{Figure Captions\markboth
        {FIGURECAPTIONS}{FIGURECAPTIONS}}\list
        {Figure \arabic{enumi}:\hfill}{\settowidth\labelwidth{Figure
999:}
        \leftmargin\labelwidth
        \advance\leftmargin\labelsep\usecounter{enumi}}}
 \relax
\def\tablecap{\section*{Table Captions\markboth
        {TABLECAPTIONS}{TABLECAPTIONS}}\list
        {Table \arabic{enumi}:\hfill}{\settowidth\labelwidth{Table
999:}
        \leftmargin\labelwidth
        \advance\leftmargin\labelsep\usecounter{enumi}}}
 \relax
\def\reflist{\section*{References\markboth
        {REFLIST}{REFLIST}}\list
        {[\arabic{enumi}]\hfill}{\settowidth\labelwidth{[999]}
        \leftmargin\labelwidth
        \advance\leftmargin\labelsep\usecounter{enumi}}}
 \relax
\makeatletter
\newcounter{pubctr}
\def\publist{\@ifnextchar[{\@publist}{\@@publist}}
\def\@publist[#1]{\list
        {[\arabic{pubctr}]\hfill}{\settowidth\labelwidth{[999]}
        \leftmargin\labelwidth
        \advance\leftmargin\labelsep
        \@nmbrlisttrue\def\@listctr{pubctr}
        \setcounter{pubctr}{#1}\addtocounter{pubctr}{-1}}}
\def\@@publist{\list
        {[\arabic{pubctr}]\hfill}{\settowidth\labelwidth{[999]}
        \leftmargin\labelwidth
        \advance\leftmargin\labelsep
        \@nmbrlisttrue\def\@listctr{pubctr}}}
 \relax
\makeatother
\newskip\humongous \humongous=0pt plus 1000pt minus 1000pt

\newif\ifdtup

\relax

\def\be{\begin{equation}}
\def\ee{\end{equation}}
\def\ba{\begin{eqnarray}}
\def\ea{\end{eqnarray}}

\def\del{\partial}

\def\r{\rho}

\def\b{\beta}

\def\d{\delta}

\def\m{\mu}
\def\n{\nu}
\def\om{\omega}
\def\Om{\Omega}
\def\l{\lambda}
\def\L{\Lambda}
\def\s{\sigma}

\def\cL{{\cal L}}

\def\no{\noindent}

\def\qq{\qquad}

\def\IR{\relax{\rm I\kern-.18em R}}

\def \ha {{1\over 2}}

\def \ov {\over}

\def\IR{\relax{\rm I\kern-.18em R}}
\def\IL{\relax{\rm I\kern-.18em L}}

\def\inv{^{\raise.15ex\hbox{${\scriptscriptstyle -}$}\kern-.05em 1}}

\def\cL{{\cal L}}

\def\Tr{{\rm Tr}}

\begin{document}

\renewcommand{\theequation}{\thesection.\arabic{equation}}
\csname @addtoreset\endcsname{equation}{section}

\newcommand{\beq}{\begin{equation}}
\newcommand{\eeq}[1]{\label{#1}\end{equation}}
\newcommand{\ber}{\begin{equation}}
\newcommand{\eer}[1]{\label{#1}\end{equation}}
\newcommand{\eqn}[1]{(\ref{#1})}
\begin{titlepage}
\begin{center}

\hfill CERN-TH-2017-148

${}$
\vskip .2 in

{\large\bf Integrable flows between exact CFTs}

\vskip 0.4in

{\bf George Georgiou$^1$}\ and \ {\bf Konstantinos Sfetsos}$^{2,3}$
\vskip 0.1in

\vskip 0.1in
{\em
${}^1$Institute of Nuclear and Particle Physics,
\\ National Center for Scientific Research Demokritos,
\\
Ag. Paraskevi, GR-15310 Athens, Greece
}
\vskip 0.1in

 {\em
${}^2$Department of Nuclear and Particle Physics,\\
Faculty of Physics, National and Kapodistrian University of Athens,\\
Athens 15784, Greece\\
}
\vskip 0.1in

 {\em
${}^3$Theoretical Physics Department, \\
CERN, CH-1211 Geneva 23, Switzerland\\
}

\vskip 0.1in

{\footnotesize \texttt georgiou@inp.demokritos.gr, ksfetsos@phys.uoa.gr}


\vskip .5in
\end{center}

\centerline{\bf Abstract}

\no
We explicitly construct families of integrable $\s$-model actions smoothly interpolating between exact CFTs.
In the ultraviolet the theory is the direct product of two current algebras at levels $k_1$ and $k_2$. In the  infrared
and for the case of two deformation matrices the CFT involves  a coset CFT, whereas for a single matrix deformation it is given by the ultraviolet direct product theories but at levels $k_1$ and $k_2-k_1$.
For isotropic deformations we demonstrate integrability.
In this case we also compute the exact beta-function for the deformation parameters using gravitational methods. This is shown to coincide with previous results obtained using perturbation theory and non-perturbative symmetries.

\vskip .4in
\noindent
\end{titlepage}
\vfill
\eject

\newpage

\tableofcontents

\noindent

\def\baselinestretch{1.2}
\baselineskip 20 pt
\noindent


\setcounter{equation}{0}
\section{Introduction }

In recent years all-loop effective actions describing various deformations of  current algebra conformal field theories (CFTs) have been constructed. These
deformed theories posess novel non-perturbative in the deformation parameters quantum symmetries which are classically realized by the effective actions.
They also provide new integrable $\s$-models and serve as the starting point for constructing new type-II supergravity solutions.

\no
The prototype example for these developments has been a deformation of the WZW action $S_k(g)$, where
$g$ is an element of a general semi-simple group $G$. The  WZW action perturbed by a current bilinear is given by
\be
\label{pert1}
S = S_{k}(g) + {k\ov \pi} \int d^2 \s \l^{ab} J_+^a J_-^b \ .
\ee

In the absence of the bilinear term,  the $J_+^a$'s and the $J_-^b$'s obey two commuting current algebras both at the same positive level $k$.
The coupling constants $\l_{ab}$, where $a,b=1,2,\dots, \dim G$, are elements of a matrix $\l$.
Due to the deformation, the original WZW model is driven towards the infrared (IR) away from the conformal point which is in the ultraviolet (UV).

\no
The effective action for \eqn{pert1} which takes into account all loops in $\l$ but is valid
for large $k$, was constructed in \cite{Sfetsos:2013wia} and for $\l_{ab}=\l \d_{ab}$
was shown to correspond to an integrable $\s$-model.
These models are generically called $\l$-deformed.
These considerations were extended to the case where the WZW model is replaced by a
coset CFT \cite{Sfetsos:2013wia,Hollowood:2014rla,Hollowood:2014qma}, as well as to the case of supergroups \cite{Hollowood:2014rla,Hollowood:2014qma}. The computation of the renormalization group (RG) flow equations that are exact in $\l$ but
for large $k$ using gravitational methods was performed in
\cite{Itsios:2014lca,Sfetsos:2014jfa}. The results agree with those obtained from field theoretical  methods in the past \cite{Kutasov:1989dt, Gerganov:2000mt} and more recently in \cite{Appadu:2015nfa}. Furthermore, deformed models of low dimensionality have been embedded to supergravity \cite{Sfetsos:2014cea,Demulder:2015lva,
Borsato:2016zcf,Chervonyi:2016ajp,Chervonyi:2016bfl}.
Integrability has also been shown for $\l$-deformed models corresponding to symmetric coset spaces \cite{Hollowood:2014rla,Hollowood:2014qma}, for the $SU(2)$ group case and diagonal matrix $\l$ \cite{Sfetsos:2014lla},
and for the models in \cite{Sfetsos:2015nya}. Moreover,
the $\l$-deformations were shown to be related via Poisson-Lie T-duality \cite{KS95a} and appropriate analytic continuations  \cite{Vicedo:2015pna,Hoare:2015gda}, \cite{Sfetsos:2015nya,Klimcik:2015gba,Klimcik:2016rov} to a different type of integrable deformations,
the so-called $\eta$-deformations for group and coset spaces introduced in \cite{Klimcik:2002zj,Klimcik:2008eq,Klimcik:2014}
and \cite{Delduc:2013fga,Delduc:2013qra,Arutyunov:2013ega}, respectively.
For the case of isotropic couplings, i.e. $\l_{ab}=\l \d_{ab}$ and for $k\gg 1$, all-loop correlators of current and primary field operators have been computed in \cite{Georgiou:2016iom, Georgiou:2015nka}. 
In these computations a few terms obtained using perturbation theory and the non-perturbative symmetry, argued via path integral considerations in \cite{Kutasov:1989aw},
were enough to obtain the exact results.
Other selected and related recent works can be found in \cite{selected}. 

 One may wonder in which sense the  action constructed in \cite{Sfetsos:2013wia} is the unique all-loop in $\l$ effective action for \eqn{pert1}. 
There are several facts pointing towards that intepretation.
First of all, both actions have the same global symmetries and coincide up to ${\cal O}(\l)$. Moreover,  the action of \cite{Sfetsos:2013wia} realizes classically the
quantum symmetries of \eqn{pert1} and it correctly reproduces the all-loop beta-functions and anomalous dimensions of the currents. 
Finally, the exact in $\l$ operator product expansion (OPE) of the currents was computed  and the corresponding Poisson brackets were extracted 
 \cite{Georgiou:2015nka} (see section 6). These were shown to coincide
with the bracket algebra of \cite{Balog:1993es} which is realized by the action of \cite{Sfetsos:2013wia}.

\no
The above developments make it apparent that it is worth to pursue further this
line of research. Consider the following modification of \eqn{pert1}
\be
\label{pert2}
S =S_{CFT} + {\sqrt{k_1k_2} \ov \pi} \int d^2 \s \l^{ab} J_+^a J_-^b \ ,
\ee
 where $S_{CFT} $ is the action of a CFT possessing left and right conserved currents $J_+^a$ and $J_-^a$ which obey the standard Kac-Moody algebras with levels $k_1$ and $k_2$, respectively. 
This theory was studied in \cite{LeClair:2001yp} where the beta-functions of the model were evaluated. Subsequently, the authors of
\cite{Georgiou:2016zyo}  computed the exact anomalous dimensions of current and
primary field operators using CFT perturbation theory and non-perturbative  quantum symmetries argued in \cite{Kutasov:1989aw}.
The essential feature of the model in \eqref{pert2} is that under 
the RG flow a new fixed point in the IR is reached. This fixed point is not present in the case where the levels are equal. This feature is very appealing but at the same time the theory is chiral due to the levels being unequal. This fact makes the Lagrangian description of the theory, not to mention the construction of an effective all loop action, an important and highly non-trivial task which remained elusive until the present paper.

\no
In a parallel development the all-loop effective action of two WZW actions for the group elements $g_1,g_2\in G$ mutually interacting via current bilinears, i.e.
\be
\label{pert3}
S = S_{k}(g_1) + S_{k}(g_2) + {k\ov \pi} \int d^2 \s
\left(  (\l_1)^{ab} J_{1+}^a J_{2-}^b + (\l_2)^{ab} J_{2+}^a J_{1-}^b\right) \ ,
\ee
was constructed  in \cite{Georgiou:2016urf}. In this model there are four commuting current algebras generated by the $J_{1+}^a$'s, $J_{2+}^a$'s as well by the $J_{1-}^a$'s, $J_{2-}^a$'s, all at level
$k$. The current bilinear terms above represent mutual interactions between
the two WZW  models. Self-interacting terms of the form appearing in \eqn{pert1} and
\eqn{pert2} are absent. The anomalous dimensions of current and primaries in this theory were computed using CFT techniques and symmetry arguments in \cite{Georgiou:2017aei}.
It turns out that the effective action corresponding to \eqn{pert3} is canonically equivalent \cite{Georgiou:2017oly} to the effective action of the sum of two models of the form \eqn{pert1}. 
As such the beta-functions for
the couplings are identical and the anomalous dimensions of operators are related.

\no
The purpose of the present paper is to find an action realizing all loop
effects  of the theory \eqn{pert2}.
We will show that this will be provided by a modification
of the procedure in \cite{Georgiou:2016urf} that led to the effective action for
\eqn{pert3}.
\renewcommand{\theequation}{\thesection.\arabic{equation}}
Our construction utilizes two group elements of a general semisimple group
and may have two or one distinct deformation matrices.
The $\s$-models that we will construct are integrable and smoothly interpolate between exact CFTs.
At the UV point the theory is described by the sum of two WZW models one  at level $k_1$ and the other at level $k_2$. As soon as the perturbation is turned on our models are driven towards another fixed point in the IR.
When both couplings are non-zero the IR  CFT is described by a coset CFT the precise nature of which will be analysed in section 5.
When one of the couplings has been set to zero
the IR  CFT is given  by the sum of two WZW models one  at level $k_1$ and the other at level $k_2-k_1$.
In both cases the flow respects Zamolodchikov's $c$-theorem.
For isotropic deformations we demonstrate that the theory is integrable. We explicitly construct the Lax pairs and show that the conserved charges are in involution.
We then proceed to compute the exact beta-function for the deformation parameters using gravitational methods. This is shown to coincide with previous results obtained using perturbation theory and non-perturbative symmetries of the theory \cite{LeClair:2001yp,Georgiou:2016zyo}. Finally, we present our conclusions as well as some future research directions.

\section{Constructing the Lagrangian}

In order to make the line of reasoning transparent, we first briefly review the integrable models constructed in \cite{Georgiou:2016urf} since it is a modification of these models that will give the Lagrangian we are after.  Nevertheless, the reader may skip this part and jump directly to the proposed action \eqref{gaufix} which we subsequently  actually use.
The basic idea was to generalize the construction of the
$\l$-deformed models of \cite{Sfetsos:2013wia}, by first replacing  the usual gauged WZW action by the following left-right asymmetric gauged action for a general semisimple group $G$ \cite{Witten:1991mm}
\be
\label{acct1}
\begin{split}
&
S_k(g,A_\pm,B_\pm) = S_{k}(g)
 +{k\ov \pi} \int d^2\s \ \Tr \big(A_- \del_+ g g^{-1} - B_+ g^{-1} \del_- g
+ A_- g B_+ g^{-1}
\\
&\qq\qq \qq\qq  -\ha  A_- A_+ - \ha B_+ B_- \big)\ ,
\end{split}
\ee
where $S_k(g)$ is the WZW action for the group element $g\in G$.  Note also
the use of two different gauge fields $A_\pm $ and $B_\pm$ taking values in the
corresponding Lie algebra.
Under the infinitesimal gauge transformations
\be
  \d g = g u_R - u_L g \ ,\qq
\d A_\pm =-\del_\pm u_L + [A_\pm, u_L]\ ,\qq
\d B_\pm =-\del_\pm u_R + [B_\pm, u_R]\ ,
\ee
which have different infinitesimal parameters for the left and the right transformations,
the action \eqn{acct1} changes as
\be
\d S_k(g,A_\pm,B_\pm)  = {k\ov 2\pi} \int d^2\s \
\Tr\big[ (A_+ \del_- u_L - A_- \del_+ u_L) - (B_+ \del_- u_R - B_- \del_+ u_R)\big]\ .
\label{anogwzw}
\ee
This is independent of the group element.
The strategy of \cite{Georgiou:2016urf} was to combine two of the aforementioned actions with two gauged PCMs as follows
\be
\begin{split}
& S_{k}(g_1,g_2, \tilde g_1,\tilde g_2,A_\pm, B_\pm)= S_{k}(g_1,A_\pm,B_\pm)
+ S_{k}(g_2,B_\pm,A_\pm)
\label{fgje3}
\\
& \qq \qq\quad -{1 \ov \pi}   \int d^2\s \ \Tr \big(t^a \tilde  g_1^{-1} D_+ \tilde g_1)E_{1ab} \Tr(t^b \tilde g_1^{-1} D_- \tilde g_1\big)
\\
&\qq \qq\quad -{1 \ov \pi}   \int  d^2\s \
\Tr \big(t^a\tilde  g_2^{-1} D_+ \tilde g_2)E_{2ab} \Tr(t^b \tilde g_2^{-1} D_- \tilde g_2\big)\ .
\end{split}
\ee
Note that the role of $A_\pm$ and  $B_\pm$ is exchanged in the two gauged WZW
actions. The covariant derivatives acting on the group elements defining the PCMs are
$ D_\pm \tilde g_1 = \del_\pm \tilde g_1 - A_\pm \tilde g_1$ and
$ D_\pm \tilde g_2 = \del_\pm \tilde g_2 - B_\pm   \tilde g_2 $. The matrices $E_1$ and $E_2$ parametrize the corresponding couplings.

\no
The virtue of this action is that it is invariant under the set of transformations
\ba
&&  \d g_1 =g_1 u_R - u_L g_1 \ ,\qq \d g_2 =  g_2 u_L - u_R g_2\ ,
\nonumber\\
&& \d \tilde g_1 =- u_L \tilde g_1 \ ,\qq \d g_2 =  - u_R \tilde g_2\ ,
\\
&&
\d A_\pm =-\del_\pm u_L + [A_\pm, u_L]\ ,\qq
\d B_\pm =-\del_\pm u_R + [B_\pm, u_R]\ .
\nonumber
\ea
Indeed, in the first line the variation of the first term in \eqn{fgje3} cancels that of the second term. The second and third lines involving the PCMs are invariant by themselves.
The next step taken was to completely fix  the gauge by choosing
$\tilde g_1=\tilde g_2=\mathbb{I}$. The resulting gauge fixed action was  given by
\be
\begin{split}
&  S(g_1,g_2, A_\pm, B_\pm) = S_{k_1}(g_1) + S_{k_2}(g_2)- {\sqrt{k_1 k_2}\ov \pi} \int d^2\s\ \Tr\big(A_+ \l_1^{-1} A_-  +B_+ \l_2^{-1} B_- \big)
\\
&\qq\qq +{k_1\ov \pi} \int d^2\s \ \Tr \big(A_- \del_+ g_1 g_1^{-1}   - B_+ g_1^{-1} \del_- g_1+ A_- g_1 B_+ g_1^{-1}\big) \\
&\qq\qq +{k_2\ov \pi} \int d^2\s \ \Tr \big(B_- \del_+ g_2 g_2^{-1}   - A_+ g_2^{-1} \del_- g_2+ B_- g_2 A_+ g_2^{-1} \big) \ ,
\label{gaufix}
 \end{split}
\ee
where we have introduced the  parameters
\be
\l_i  = \sqrt{k_1 k_2}\ (k\mathbb{I} + E_i )^{-1} \ ,\quad i=1,2\ ,\qquad k = {k_1+k_2\ov 2}\ .
\ee
To be precise the action obtained in \cite{Georgiou:2016urf} is the one presented above but with $k_1=k_2$.
Nevertheless, in \eqref{gaufix} we have relaxed the condition that the two asymmetrically gauged WZW models must have the same level. We postulate this action to be our starting point.
In what follows we will see that this modification drastically changes the all-loop $\b$-functions of the model which acquire new fixed points in the IR under the flow of the renormalization group.
This fact will be verified by using the gravity background  generated by the all-loop effective action \eqref{gaufix} which is obtained after integrating out the gauge fields $A_{\pm}$ and $B_{\pm}$. Our results will be in complete agreement with results obtained employing CFT methods and symmetry considerations in \cite{LeClair:2001yp} and in \cite{Georgiou:2016zyo}.

\no
Integrating out the gauge fields in the action \eqn{gaufix} we find that
\be
\label{abp}
\begin{split}
&
A_+ = i (\mathbb{I}-\l_1^T D_1 \l_2^T D_2)^{-1}\l_1^T ( \l_0 J_{1+} +  D_1\l_2^T J_{2+})\ ,
\\
&
A_- = -i (\mathbb{I}-\l_1 D_2^T \l_2 D_1^T)^{-1}\l_1 ( \l_0^{-1} J_{2-} + D^T_2\l_2 J_{1-})
\end{split}
\ee
and that
\be
\label{abm}
\begin{split}
& B_+ = i (\mathbb{I}-\l_2^T D_2 \l_1^T D_1)^{-1}\l_2^T ( \l_0^{-1}J_{2+} +  D_2\l_1^T J_{1+})\ ,
\\
&
B_- = -i (\mathbb{I}-\l_2 D_1^T \l_1 D_2^T)^{-1} \l_2 (  \l_0 J_{1-} +  D^{T}_1\l_1 J_{2-})\ .
\end{split}
\ee
The matrices $D_{ab}$ and the currents $J^a_{\pm}$ are defined as
\be
\label{hg3}
J^a_+ = - i\, \Tr(t^a \del_+ g g^{-1}) ,\qq J^a_- = - i\, \Tr(t^a g^{-1}\del_- g )\ ,
\qq D_{ab}= \Tr(t_a g t_b g^{-1})\  ,
\ee
where the $t^a$'s are Hermitian matrices.
When a current or the matrix $D$ has an index $1$ or $2$ this means that one should use the corresponding group element in its definition.
In addition, we have defined the ratio of the two levels
\be
\label{levl0}
 \l_0=\sqrt{{k_1 \ov k_2}}\ ,
\ee
which with no loss of generality can be taken to be less than one.

\no
The substitution of the expressions for the gauge fields into the action results in a
$\s$-model action which can be written in matrix notation as
\be
\begin{split}
&  S_{k_1,k_2,\l_1,\l_2}(g_1,g_2) = S_{k_1}(g_1) + S_{k_2}(g_2)
\\
&\qq\quad + {1\ov \pi} \int  d^2\s  \left(\!\! \begin{array}{cc}
    J_{1+}\! &\! J_{2+} \end{array}\!\!  \right)
\left(  \begin{array}{cc}
     k_1 \L_{21}\l_1 D_2^T\l_2 &  k_2 \l_0 \L_{21}\l_1 \\
   k_1 \l_0^{-1} \L_{12}\l_2 &k_2 \L_{12} \l_2 D_1^T\l_1\\
  \end{array} \right)
  \left(\!\! \begin{array}{c}
    J_{1-} \\ J_{2-} \end{array}\!\!\! \right) \ , 
\end{split}
\label{defactigen}
\ee
where we have also defined the matrices
\be
\L_{12}= (\mathbb{I} - \l_2 D_1^T \l_1 D_2^T)^{-1}\ ,\qq
\L_{21}= (\mathbb{I} - \l_1 D_2^T \l_2 D_1^T)^{-1}\ .
\ee
The above action is by construction symmetric under the exchange of
the two original models, i.e. indices $1$ and $2$.
More importantly, the model with equal levels constructed in \cite{Georgiou:2016urf} inherits a remarkable duality-type symmetry to the present model \eqref{defactigen}. This symmetry reads
\be
 k_1 \to -k_2 \,  \quad k_2 \to -k_1 \ , \quad \l_1 \to  \l_1^{-1} \ ,\quad  \l_2 \to  \l_2^{-1}\ ,
 \quad g_1\to g_2^{-1}\quad g_2\to g_1^{-1}\  .
\label{symmdual}
\ee
The proof uses the fact that under \eqn{symmdual}
\be
\begin{split}
&
D_1\to D_2^T\ ,\qquad J_{1+}\to - D_2^T J_{2+}\ ,\qq J_{1-}\to - D_2 J_{2-}\  ,
\\
&
D_1\to D_2^T\ ,\qquad  J_{2+}\to - D_1^T J_{1+}\ , \qq J_{2-}\to -  D_1 J_{1-}\  .
\end{split}
\ee
The action \eqn{defactigen} may have  additional global isometries for specific choices of the deformation matrices.
In particular, if the $\l_i$'s are proportional to the identity the action has the global symmetry
\be
g_1\to \L_L^{-1} g_1 \L_R \ ,\qq   g_2\to \L_R^{-1} g_2 \L_L\ , \qq \L_L,\L_R\in G\ .
\ee
For general deformation matrices this symmetry is partially or completely broken.

\no
For small elements of the matrices $\l_i$'s the action \eqn{defactigen} can be approximated by
\be
\label{expan1}
S_{k_1,k_2,\l_1,\l_2}(g_1,g_2) = S_{k_1}(g_1) + S_{k_2}(g_2) + {\sqrt{k_1 k_2}\ov \pi}
\int d^2\s\ (\l_1^{ab} J^a_{1+} J^b_{2-} + \l_2^{ab} J^a_{2+} J^b_{1-}) + \cdots \ .
\ee
It represents a current-current interaction of the two original WZW actions, one at level $k_1$ and the other at level $k_2$.  In fact, the action \eqn{defactigen} can be considered as the effective action for the theory \eqref{expan1}
that incorporates all-loop effects in the deformation parameter $\l_i$.

At this point let us argue that although the theory \eqref{expan1} is left-right symmetric it reproduces correctly the all-loop correlation functions of the chiral model of \eqref{pert2}. The argument goes as the one in \cite{Georgiou:2017aei}.
All correlation functions involving the operators of the set {${\mathcal O}=\{J^a_{1+},\,\, J^a_{2-}, \,\,J^a_{1+}  J^b_{2-},\,\,\cdots\}$} or any other composite operator built from them can be calculated as if the vertex proportional to $\l_2$ in \eqref{expan1} was absent. This is so because the OPE of the currents appearing in the first interaction vertex of \eqref{expan1} with any of the currents appearing in the second interaction vertex of \eqref{expan1} is regular. This means that if one restricts himself to the set of operators ${\mathcal O}$ it is as if he effectively sets $\l_2$ to zero. Thus, not only the $\b$-functions but also all current correlation functions of the models \eqref{expan1} and \eqref{pert2} coincide to all-orders in the $\l$-,
as well as in the $k$-expansion.

\subsection{Two limits}

\subsubsection{One vanishing deformation matrix}

Let one of the deformation parameters in \eqn{defactigen} approach zero, 
say $\l_2\to 0$ and rename $\l_1$ to $\l$.
Then the action \eqn{defactigen} simplifies drastically to the following form
\be
S_{k_1,k_2,\l}(g_1,g_2) = S_{k_1}(g_1) + S_{k_2}(g_2) + {\sqrt{k_1 k_2}\ov \pi} \int d^2\s
\l_{ab} J_{1+}^a J_{2-}^b\ ,
\label{accctl2o}
\ee
making the perturbative expression \eqn{expan1} exact. Note that this special case, but with $k_1=k_2$, has been examined before in \cite{Solovev:1993he,Hull:1995gj}.
Since the currents $J_{2+}^a $ and $J_{1-}^a $ do not appear in the action they do not acquire anomalous dimensions.
This fact implies that \eqn{accctl2o} should have on-shell chiral and anti-chiral currents.
Following a procedure parallel to that in \cite{Georgiou:2017aei} we find that the equations of motion from varying the groups elements can be cast in the form
\be
\label{eqqq2}
\begin{split}
& \del_- {\cal J}_+ = 0 \ ,\qq {\cal J}_+ =   \l_0^{-1} J_{2+}  + D_2 \l^T J_{1+}\ ,
\\
& \del_+ {\cal J}_- = 0 \ ,\qq {\cal J}_- = \l_0 J_{1-}  + D_1^T \l J_{2-}\ .
\end{split}
\ee
To prove the above equations we have used the identities $(D^T \del_- D)^{ab} = f^{ab}{}_c J^c_-$ and $(\del_+ D D^T)^{ab} = f^{ab}{}_c J^c_+$.
The above chiral and anti-chiral conserved currents ${\cal J}_\pm$ are
deformations of $J_{2+}$ and $J_{1-}$
to which they reduce for $\l=0$. This is consistent with their
vanishing anomalous dimensions.

\subsubsection{Zooming in}

A second interesting limiting case involves taking  one the group elements to unity and at the same time the corresponding
level to infinity. Specifically, choosing $g_2$ as the relevant group element, we have that
\be\label{limitt}
g_2= \mathbb{I} + i\l_0  {v_a t^a} + \dots\ ,\qquad k_2\to \infty\ .
\ee
Hence in this limit the parameter $\l_0\to 0 $.
We also drop the subsctript from $g_1$ and the level $k_1$.
In this limit the action \eqn{defactigen} simplifies to
\be
\begin{split}
&  S_{k,\l_1,\l_2}(g,v) = S_k(g) + {k\ov 2\pi} \int d^2\s \del_+ v^a \del_- v^a
\\
&\qq\quad + {k\ov \pi} \int  d^2\s  \left(\!\! \begin{array}{cc}
    J_{+}\! &\! \del_+ v \end{array}\!\!  \right)
\left(  \begin{array}{cc}
     \tilde \L_{21}\l_1 \l_2 &   \tilde \L_{21}\l_1 \\
   \tilde \L_{12}\l_2 & \tilde \L_{12} \l_2 D^T\l_1\\
  \end{array} \right)
  \left(\!\! \begin{array}{c}
    J_{-} \\ \del_-v \end{array}\!\!\! \right),
\end{split}
\label{defactigen2}
\ee
where
\be
\tilde \L_{12}= (\mathbb{I} - \l_2 D^T \l_1 )^{-1}\ ,\qq
\tilde \L_{21}= (\mathbb{I} - \l_1 \l_2 D^T)^{-1}\ .
\ee
 Note that the limit \eqref{limitt}, makes the current algebras generated by $J_{2+} $ and  $J_{2-} $ Abelian. Therefore, the action \eqref{defactigen2} represents the effective action for the mutual interaction of a WZW model at level $k$ with an Abelian theory of equal dimensionality.

\section{Integrability}

In this section, we prove that the $\s$-model action \eqn{defactigen}
is integrable when the matrices $\l_1$ and $\l_2$ are proportional to the identity, that is when $(\l_1)_{ab}=\l_1 \d_{ab}$ and $(\l_2)_{ab}=\l_2\d_{ab} $.
For the case of equal levels integrability has been shown in \cite{Georgiou:2016urf,Georgiou:2017oly}.
It is remarkable that it is preserved for unequal levels as well.

\subsection{Lax pairs and charges in involution}

The integrability of \eqn{defactigen}  is  more conveniently examined if one chooses to
work with the action \eqn{gaufix} before integrating out the gauge fields.

\no
Varying \eqn{gaufix} with respect
to $B_\pm$ and $A_\pm$ we find the following constraints
\be
\qq D_+ g_1\, g_1^{-1} =  (\l_0^{-1}\l_1^{-T}-1) A_+ \ ,\qq
g_2^{-1} D_- g_2 = - (\l_0 \l_1^{-1}-1) A_-
\label{dggd}
\ee
and
\be
\qq D_+ g_2\, g_2^{-1} =  (\l_0\l_2^{-T}-1) B_+ \ ,\qq
g_1^{-1} D_- g_1 = - (\l_0^{-1}\l_2^{-1}-1) B_- \ ,
\label{dggd2}
\ee
respectively. Varying the action with respect to group elements $g_1$ and $g_2$ results into
\be
\label{eqg1g2}
D_ -(D_+ g_1 g_1^{-1})= F_{+-}^{(A)}\ ,\qq D_ -(D_+ g_2 g_2^{-1})
= F_{+-}^{(B)}\ ,
\ee
where
\be
F_{+-}^{(A)}=\del_+ A_- - \del_- A_+ - [A_+,A_-]\ ,\qq
F_{+-}^{(B)}=\del_+ B_- - \del_- B_+ - [B_+,B_-]\ .
\ee
Equivalently, equations \eqref{eqg1g2} can be written as
\be
D_+(g_1^{-1}D_- g_1)= F_{+-}^{(B)}\ ,\qq
D_+(g_2^{-1}D_- g_2)= F_{+-}^{(A)}\ .
\label{eqg1g22}
\ee
The definitions of the covariant derivatives depend on the transformation properties of
the object on which they act. For example, the action on the group element $g_1$ involves both the $A_\pm$ and the $B_\pm$ gauge fields, that is $D_\pm g_1= \del_\pm g_1 -A_\pm g_1 + g_1B_\pm$, while
$D_- (D_+ g_1  g_1^{-1})= \del_-(D_+ g_1  g_1^{-1}) -[A_-,(D_+ g_1  g_1^{-1})]$
involves only the $A_\pm$.
The next step is to substitute the constraint equations in \eqn{eqg1g2} and \eqn{eqg1g22}. After some algebra one obtains that
\be
\begin{split}
\label{eomAinitial1}
&\del_+ A_- - \l_0^{-1}\l_1^{-T} \del_-  A_+ = \l_0^{-1} [\l_1^{-T} A_+,A_-]\ ,
\\
&\l_0 \l_1^{-1}\del_+A_--\del_-A_+= \l_0 [A_+,\l_1^{-1}A_-]\
\end{split}
\ee
and that
\be
\begin{split}
\label{eomAinitial2}
&  \del_+ B_-  - \l_0\l_2^{-T}\del_-  B_+ = \l_0 [\l_2^{-T} B_+,B_-]\ ,
\\
&
\l_0^{-1}\l_2^{-1}\del_+ B_-- \del_-B_+= \l_0^{-1} [B_+,\l_2^{-1}B_-]\ .
\end{split}
\ee
We conclude that the equations of motion seemingly decouple forming two independent sets.
Nevertheless, the fields $A_\pm$ and $B_\pm$ depend on both $(g_1,\l_1)$ and $(g_2,\l_2)$.

For the special case where $(\l_1)_{ab}=\l_1 \d_{ab}$ and $(\l_2)_{ab}=\l_2\d_{ab} $ equations
\eqn{eomAinitial1} and \eqn{eomAinitial2} can be rewritten in the form
\be
\begin{split}
\label{eomAfinal1}
 \del_-  A_+ = -{1-\l_0 \l_1 \ov 1-\l_1^2}[ A_+,A_-]\ ,
\qquad
\del_+A_-={1-\l_0^{-1} \l_1 \ov 1-\l_1^2}[A_+,A_-]\
\end{split}
\ee
and
\be
\begin{split}
\label{eomAfinal2}
 \del_-  B_+ = -{1-\l_0^{-1} \l_2 \ov 1-\l_2^2}[ B_+,B_-]\ ,
\qquad
\del_+B_-={1-\l_0\l_2 \ov 1-\l_2^2}[B_+,B_-]\ .
\end{split}
\ee

\no
We are now in a position to write down the Lax pairs which imply that the theory is integrable.  The Lax pairs should satisfy
the relations
\be
\label{Lax}
\del_+\cL^{(i)}_- -\del_-\cL^{(i)}_+=[\cL^{(i)}_+,\cL^{(i)}_-]\ ,\qq i=1,2\ ,
\ee
where each of the Lax pairs $\cL^{(i)}_\pm(\tau,\sigma;\zeta_i)$ will depend on a spectral
parameter $\zeta_i\in\mathbb{C}$.
Furthemore, for notational convenience
we drop the subscript from $\zeta_i$. One can easily show that the Lax pair in the case of \eqn{eomAfinal1} is given by
 \be
 \cL_\pm
=  {2 \zeta \ov \zeta \mp 1} \tilde A_\pm \ ,\quad \tilde A_+= {1-\l_0^{-1} \l_1 \ov 1-\l_1^2}A_+, \quad \tilde A_-= {1-\l_0 \l_1 \ov 1-\l_1^2}A_-
,\, \quad\zeta\in\mathbb{C}\ .
\label{Laxpairs1}
\ee
Similarly, for  \eqn{eomAfinal2} the Lax pair is given by
 \be
 \cL_\pm
=  {2 \zeta \ov \zeta \mp 1} \tilde B_\pm \ ,\quad \tilde B_+= {1-\l_0 \l_2 \ov 1-\l_2^2}B_+, \quad \tilde B_-= {1-\l_0^{-1} \l_2 \ov 1-\l_2^2}B_-
,\, \quad\zeta\in\mathbb{C}\ .
\label{Laxpairs2}
\ee
The careful reader may have noticed that our claim that the theory is integrable is not quite proven yet. One should in addition show that the conserved
charges obtained from the two Lax pairs above  are in involution. To this end we define dressed currents
in such a way that when these are expressed in terms of canonical variables they have the same form as the corresponding currents of a WZW model.
This idea was employed for the gauged WZW models  \cite{Bowcock} and in the present context in  \cite{Georgiou:2016urf}. In particular, we define the dressed currents as
\be
\begin{split}
\label{Affine}
&\mathcal{J}^{(1)}_{+}  =D_+ g_1 g_1^{-1} + A_+ - A_-\ ,
\quad \mathcal{J}^{(1)}_{-}  = - g_1^{-1} D_- g_1+ B_{-} - B_{+}\ ,\\
&\mathcal{J}^{(2)}_{+}  =D_+ g_2 g_2^{-1} + B_{+} - B_{-}\ ,
\quad \mathcal{J}^{(2)}_{-}  =- g_2^{-1} D_- g_2 + A_- - A_+ \ .
\end{split}
\ee
These currents obey four independent commuting copies of current algebras \cite{Georgiou:2016urf}
\be
\label{Bowcock}
\{\mathcal{J}_{\pm}^{(i)a} , \mathcal{J}_{\pm}^{(i)b}\} =
\frac{2}{k_i}\,f_{abc} \mathcal{J}_{\pm}^{(i)c} \d_{\s\s'} \pm \frac{2}{k_i} \d_{ab} \d'_{\s\s'}\,, \quad
\{\mathcal{J}_{\pm}^{(i)a} , \mathcal{J}_{\mp}^{(i)b}\} =0, \quad i=1,2\ ,
\ee
which encode the canonical structure of the theory. Using the definitions \eqn{Affine}, the constraints \eqn{dggd} and \eqn{dggd2}
can be written as
\be
\label{SthroughA}
\begin{split}
&\mathcal{J}^{(1)}_{+} =\l_0^{-1} \lambda_1^{-T} A_{+} -A_{-}\ ,
\quad  \mathcal{J}^{(1)}_{-} =\l_0^{-1}  \lambda_2^{-1} B_- -B_+\ ,\\
& \mathcal{J}^{(2)}_{+} =\l_0 \lambda_2^{-T} B_{+} - B_{-}\ ,
\quad \mathcal{J}^{(2)}_{-} = \l_0 \lambda_1^{-1} A_- - A_+\ .
\end{split}
\ee
These equations can be easily inverted in order to express the $A_\pm$ and $B_\pm$
in terms of the dressed currents.
By just inspecting \eqn{SthroughA} it is easy to see that $A_\pm$ will depend only on $\mathcal{J}^{(1)}_{+}$ and $\mathcal{J}^{(2)}_{-}$
while $B_\pm$ will depend only on $\mathcal{J}^{(1)}_{-}$ and $\mathcal{J}^{(2)}_{+}$. Since the Poisson brackets of any of the variables in the set $\{\mathcal{J}^{(1)}_{+},\mathcal{J}^{(2)}_{-}\}$ with any of the variables in the set $\{\mathcal{J}^{(1)}_{-},\mathcal{J}^{(2)}_{+}\}$ is zero we conclude that the Poisson bracket between $A_{\pm}$ and $B_{\pm}$ is zero, that is
$\{A_\pm, B_\pm\}_{\rm P.B.}=0$.
This completes the proof that the charges generated by the Lax pair of \eqn{Laxpairs1} and those generated by the Lax pair of \eqn{Laxpairs2} are in involution. Thus, the theory defined by \eqn{defactigen} is integrable.\footnote{
One might wonder if the conserved changes provided by each one of the Lax pairs are among themselves in involution due
the non-ultralocal term proportional to $\d'$ in \eqn{Bowcock}. Such terms give rise to non-ultralocal terms in the Poisson algebra of the ${\cal L}_\s$ 
which is used to define the monodromy matrix and from that to construct the infinite tower of
conserved changes. Nevertheless, it has been shown in \cite{Maillet,Maillet2} that the presence of such terms does not spoil the fact that the infinite number of conserved charges are in involution (eqs. (3.23) and (3.24) of \cite{Maillet}) provided that the Poisson brackets of ${\cal L}_\s$ assume the Maillet form and the modified Yang--Baxter equation is satisfied.
Our effective action \eqref{defactigen} implies a canonical structure which is precisely a double copy of the two-parameter deformation of the PCM's canonical structure presented in \cite{Itsios:2014vfa}. In this work it was shown that the Maillet brackets are satisfied
and an explicit solution to the modified classical Yang-Baxter equation was found (see section 3, where the parameter $\r$ in there is related to the level asymmetry as in eq. (4.4) of \cite{Georgiou:2016zyo}).}
It would be interesting to see if there are other choices of the deformation matrices $\l_1$ and $\l_2$ for which the theory \eqn{defactigen} remains integrable. For the case of equal levels a classification of the different integrable cases was performed in \cite{Georgiou:2016urf} based on previous work for single $\l$-deformations  \cite{Sfetsos:2013wia,Hollowood:2014rla,Sfetsos:2014lla,Sfetsos:2015nya}.


\section{The $\beta$-function}

One may attempt to use the $\s$-model background fields for \eqn{defactigen} in order to compute using
the renormalization group equations the beta-function equations for the parameters $\l_i$, $i=1,2$. This
seems clearly a formidable task. However,  the fact that in  \eqn{expan1}
the defining CFT theories, that is the two WZW models, are decoupled, implies that there is no mixing between the two deformation parameters. The arguments in favour of that are identical to
those presented for the equal level case in \cite{Georgiou:2017aei}.
In fact the beta-function found by CFT perturbative methods in \cite{LeClair:2001yp} and in \cite{Georgiou:2016zyo} is
\be
{d\l \ov dt} = -{c_G \ov 2 \sqrt{k_1 k_2}} {\l^2 (\l-\l_0 ) (\l-\l_0^{-1})\ov (1-\l^2)^2}\ ,
\label{resubeta}
\ee
where $t=\ln \m^2$ with $\m$ an energy scale and where $\l$ could be either $\l_1$ or $\l_2$. The above formula is valid for $k_1,k_2\gg 1$.
Clearly, the RG flow is between $\l=0$ at the UV and the fixed point in the IR at $\l=\l_0$.
The other fixed point at $\l_0^{-1}$ is unphysical since via the duality \eqn{symmdual}
it corresponds to a theory with negative levels.
We also note that, even though we do not
know the RG flows equations for general $\l_{ab}$, it is guaranteed that $\l_{ab}=\l_0
\d_{ab}$ is an IR fixed point.

\no
Clearly,
one should be able to compute the above by first setting one of these parameters to zero and then using the resulting action which is much simpler.\footnote{This can be consistently done since $\l_2=0$ is a UV fixed point of the
RG flow equations.}
For the equal level
case, this approach was taken in \cite{Georgiou:2017aei} resulting into a complete agreement with  the  CFT results.
In our case we will use the action  \eqn{accctl2o} in order to compute the one loop RG flow equations \cite{honer,Friedan:1980jf,Curtright:1984dz}
\be\label{beta-gr}
{d\ov d t} (G_{\m\n}+B_{\m\n}) = R^-_{\m\n}\ ,
\ee
where the Ricci tensor includes the torsion.
In this paper, we will do this exercise for the case of isotropic couplings $\l_{ab}=\l \d_{ab}$.

The first step is to define the frames by writing the metric in the form
\be
ds^2 = {k_2\ov 2} ( e^a e^a + e^{\hat a} e^{\hat a})\ ,
\ee
where
\be
\label{framee}
e^a = \l_0\sqrt{1-\l^2}  R^a  \ ,\qq  e^{\hat a} =  L^{\hat a} + \l_0 \l R^a   \ .
\ee
We have disregarded a factor of $\displaystyle {k_2\ov 2}$ in the definition of the frames which
will be easily restored later and defined for notational convenience that
$R_1^a = R^a $ and that $L_2^a = L^{\hat a}$.

\no
We may compute all geometrical data using the relations
\be
dL^a = \ha f_{abc} L^b\wedge L^c \ ,\qq
dR^a = - \ha f_{abc} R^b\wedge R^c \ .
\label{dhj3}
\ee
We also need the antisymmetric tensor.  In a two-form notation and pulling out, as in the case for
the metric the factor $\displaystyle {k_2\ov 2}$, this is given by
\be
B =B_0 + \l_0 \l  R^a\wedge L^{\hat a}\ ,
\ee
where $B_0$ is the two-form corresponding to the two WZW models,
so that $H_0=dB_0$.
Using the frames defined in \eqn{framee} we have that
\be
\begin{split}
&
H_0 = -{1\ov 6} f_{abc}\left(\l_0^2  R^{ a}\wedge R^{ b}\wedge R^{c}+
 L^{\hat a}\wedge L^{\hat b}\wedge L^{\hat c}\right)
\\
&
\phantom{xx} =f_{abc} \Big( -{1-\l_0\l^3\ov 6 \l_0(1-\l^2)^{3/2}}  e^a \wedge e^b  \wedge e^c
 -{1\ov 6}  e^{\hat a} \wedge e^{\hat b}  \wedge e^{\hat c}
\\
&
\phantom{xxxxxxxx}
-\ha {\l^2\ov 1-\l^2} e^{\hat a }\wedge e^{ b }\wedge e^{ c }
+\ha {\l\ov \sqrt{1-\l^2}} e^{ a }\wedge e^{ \hat b }\wedge e^{ \hat c }\Big)\ .
\end{split}
\ee
In addition, the interaction term induces the following contribution to the three-form
\be
\begin{split}
& H_\l = \l_0 \l d(R^a\wedge L^{\hat a})=
-{\l_0\ov 2}\l f^{abc} R^a \wedge  L^{\hat b} \wedge (L^{\hat c} + R^c)
\\
& = {\l \ov 2\sqrt{1-\l^2}} f^{abc} \left( {\l(1-\l_0\l)\ov \l_0 (1-\l^2)} e^a\wedge e^b \wedge e^c
- e^{\hat a}\wedge e^{\hat b} \wedge e^c
+{2\l_0\l-1\ov \l_0 \sqrt{1-\l^2}}   e^{\hat a}\wedge e^{b} \wedge e^c\right)\ .
\end{split}
\ee
As a result the field strength of the $B$-field reads
\be
\begin{split}
&
H= d B =H_0 + H_\l\ = - {1\ov 6} f_{abc} \bigg({1-3 \l^2 +2 \l_0 \l^3\ov \l_0(1-\l^2)^{3/2}}
e^a\wedge e^b \wedge e^c
\\
&\qq\qq + e^{\hat a}\wedge e^{\hat b} \wedge e^{\hat c} + {3\l(1-\l_0\l)\ov \l_0(1-\l^2)} e^{\hat a}\wedge e^b \wedge e^c\bigg)\ .
\end{split}
\ee
From the last equation it is straightforward  to read off the components
\be
H_{abc} =-{1-3 \l^2 +2 \l_0 \l^3\ov \l_0(1-\l^2)^{3/2}}  f_{abc}\ ,\quad H_{\hat a\hat b\hat c} = - f_{abc}\ ,
\quad H_{\hat a bc}=- {\l(1-\l_0\l)\ov \l_0(1-\l^2)} f_{abc} \ .
\ee

\no
In a double index notation $A=(a,\hat a)$ the geometric data can be found using the relation
\be
de^A + \om^{A}{}_{B} \wedge e^B= 0 \ .
\label{spiinn}
\ee
From this ones extracts the spin connection one-form $\om^{AB}$ and finds that
\be
\begin{split}
& \om^{ab}= f_{abc}\left(\om_1 e^c + \om_2 e^{\hat c}\right)\  ,
\\
&
\om^{a\hat b}=\om^{\hat a b}=
 f_{abc}\left(\om_3 e^c + \om_4 e^{\hat c}\right)\  ,
\\
&
\om^{\hat a \hat b}=
 f_{abc}\left(\om_5 e^c + \om_6 e^{\hat c}\right)\  ,
\end{split}
\ee
where
\be
\begin{split}
& \om_1= -{1\ov 2 \l_0 \sqrt{1-\l^2}}\ ,\qq \om_2= {\l(1-\l_0 \l)\ov 2 \l_0 (1-\l^2)}\ ,
\\
& \om_3 =- {\l(1-\l_0 \l)\ov 2 \l_0 (1-\l^2)}\ ,\qq \om_4 =0 \ ,
 \\
& \om_5 = -{\l\ov \sqrt{1-\l^2}}\ ,\qq\quad\ \ \om_6 = \ha  \ .
\end{split}
\ee
It is convenient for our purposes
to use the spin connection with torsion. This is defined as
\be
\om_-^{AB} = \om^{AB}  - \ha H^{AB}{}_C e^C\ .
\ee
The components of the torsionfull  spin connection read
\be
\begin{split}
& \om_-^{ab}= f_{abc}\left(c_1 e^c + c_2 e^{\hat c}\right)\  ,
\\
&
\om_-^{\hat a\hat b}= f_{abc}\left(\hat c_1 e^c + \hat c_2 e^{\hat c}\right)\  ,
\\
&
\om_-^{ a \hat b}= \om_-^{ \hat a  b} = 0 \ ,
\end{split}
\ee
where
\be
\begin{split}
&c_1= -{\l^2 (1-\l_0\l)\ov \l_0 (1-\l^2)^{3/2}}\ ,\qq c_2= {\l (1-\l_0\l)\ov \l_0 (1-\l^2)}\ ,
\\
&
\hat c_1= - {\l \ov (1-\l^2)^{1/2}} \ ,\qq \quad  \hat c_2=1\ .
\end{split}
\ee
Then we proceed to compute the generalized Riemann tensor defined by
\be
R_-^{AB}=\ha R_-^{AB}{}_{CD} e^C \wedge e^D = d\om_-^{AB}+ \om_-^{AC}\wedge \om_-^{CB} \ .
\ee
In our case this takes the form
\be
\begin{split}
&
R_-^{ab}= d\om_-^{ab}
+ \om_-^{ac}\wedge \om_-^{cb} \ ,
\\
&
R_-^{a\hat b}  =0\ , \qq
R_-^{\hat a b}   =0\ ,
\\
&
 R_-^{\hat a\hat b}
= d\om_-^{\hat a\hat b} + \om_-^{\hat a\hat c}\wedge \om_-^{\hat c\hat b} \ .
\end{split}
\ee
As a result, the components the generalized Riemann tensor reads
\be
\begin{split}
&
R_-^{ab}{}_{de}= R_1 f_{abc} f_{cde}\ ,\qquad R_1 = 2c_1 \om_1  + 2c_2 \om_3 -c_1^2 \ ,
\\
&
R_-^{ab}{}_{d\hat e}=R_2 f_{abc} f_{cde} \ ,\qquad  R_2=c_1 \om_2  + c_1 \om_3
+ c_2\om_5 + c_2 \om_4-c_1c_2\ ,
\\
&
R_-^{ab}{}_{\hat d\hat e}=R_3  f_{abc} f_{cde}\ ,\qquad R_3=2c_1 \om_4  + 2c_2 \om_6 -c_2^2\ .
\end{split}
\ee
The components of $R_-^{\hat a\hat b}{}_{AB}$ are obtained by simply replacing $c_i$
by $\hat c_i$, leading to
\be
\begin{split}
&
R_-^{\hat a\hat b}{}_{de}= \hat R_1 f_{abc} f_{cde}\ ,
\qquad \hat R_1 = 2\hat  c_1 \om_1  + 2\hat  c_2 \om_3 -\hat  c_1^2\ ,
\\
&
R_-^{\hat  a \hat b}{}_{d\hat e}=\hat R_2 f_{abc} f_{cde}\ ,
\qquad  \hat R_2 =\hat  c_1 \om_2  + \hat c_1 \om_3 + \hat c_2\om_5 + \hat  c_2 \om_4- \hat  c_1 \hat  c_2 \ ,
\\
&
R_-^{\hat  a \hat  b}{}_{\hat d\hat e}= \hat R_3 f_{abc} f_{cde}\ ,
\qquad \hat R_3 =2 \hat  c_1 \om_4  + 2 \hat  c_2 \om_6 -\hat  c_2^2\ .
\end{split}
\ee
Using the definition of the Ricci tensor $R_-^{AB}= R_-^{AC}{}_{BC}$ we find that
\be
\label{kcf}
\begin{split}
&
R_-^{ab}= c_G \d_{ab} R_1\ ,\qq
R_-^{a\hat b}=  c_G \d_{ab} R_2\ ,
\\
&
R_-^{\hat a b}=  c_G \d_{ab} \hat R_2\ ,\qq
R_-^{\hat a\hat b}=  c_G \d_{ab} \hat R_1\ .
\end{split}
\ee

\no
Since the frame as defined in \eqn{framee} depends on $\l$, we convert to
the $R^a,L^{\hat a}$ basis.
We may change basis components using
\be
\begin{split}
&
R_-^{ab} e_+^ae_-^b + R_-^{a\hat b} e_+^ae_-^{\hat b}
+  R_-^{\hat a  b} e_+^{\hat a}e_-^{ b} + R_-^{\hat a \hat b} e_+^{\hat a}e_-^{\hat b}
\\
&\qq\qq= \tilde R_-^{ab} R_+^a R_-^b +  \tilde R_-^{\hat a\hat b} L_+^{\hat a} L_-^{\hat b}
+ \tilde R_-^{a\hat b} R_+^a L_-^{\hat b}  + \tilde R_-^{\hat a b} L_+^{\hat a} R_-^b \ ,
\end{split}
\ee
where $\pm$ as subscripts denote the corresponding light-cone versions of the frames where the exterior
derivative is replaced by the worldsheet derivatives. We have also used tilded symbols for the components in the $R^a,L^{\hat a}$ basis. Then we find that
\be
\begin{split}
& \tilde R_-^{ab}=\l_0^2 (1-\l^2) R_-^{ab}
+\l_0^2 \l \sqrt{1-\l^2}(R_-^{a\hat b} + R_-^{\hat a b})+\l_0^2 \l^2 R_-^{\hat a \hat b} \ ,
\\
&
\tilde R_-^{\hat a\hat b} =  R_-^{\hat a \hat b}\ ,
\\
&
\tilde R_-^{a\hat b} =\l_0 \sqrt{1-\l^2} R_-^{a\hat b} + \l_0 \l R_-^{\hat a \hat b}\ ,
\\
&
\tilde R_-^{\hat a b} =\l_0  \sqrt{1-\l^2} R_-^{\hat a  b} + \l_0 \l R_-^{\hat a \hat b}\ .
\end{split}
\ee
The above procedure applies for any tensor replacing the Ricci tensor $R_-^{AB}$.
By using  \eqn{kcf}, we obtain that
\be
\begin{split}
&
\tilde R_-^{ab} = c_G \l_0^2
 \d_{ab}\big((1-\l^2) R_1 + \l \sqrt{1-\l^2} (R_2+\hat R_2)+ \l^2 \hat R_1\big)\ ,
\\
&
\tilde R_-^{\hat a\hat b} = c_G \d_{ab} \hat R_1\ ,
\\
& \tilde R_-^{a\hat b} = c_G \l_0 \d_{ab} \big(\sqrt{1-\l^2} R_2+ \l \hat R_1\big)\ ,
\\
& \tilde R_-^{\hat a b} = c_G \l_0 \d_{ab} \big(\sqrt{1-\l^2} \hat R_2+ \l \hat R_1\big)\ .
\end{split}
\ee
Specifically, the various coefficients are given by
\be
\begin{split}
&
R_1= {\l^3 (1-\l_0 \l) (\l_0-\l)\ov \l_0^2 (1-\l^2)^3}\ ,
\quad R_2=- {\l^2 (1-\l_0 \l) (\l_0-\l)\ov \l_0^2 (1-\l^2)^{5/2}}\ ,
\\
& R_3= {\l (1-\l_0 \l) (\l_0-\l)\ov \l_0^2 (1-\l^2)^2}\ ,\qq
\hat  R_1 = \hat  R_2=\hat  R_3 = 0 \ ,
\end{split}
\ee
from which we find that
\be
\tilde R_-^{a\hat b}=  -{c_G}\d_{ab} {\l^2 (\l-\l_0 ) (\l-\l_0^{-1})\ov (1-\l^2)^2}\ .
\ee
All other components of the Ricci tensor vanish. After restoring the overall factor of $\displaystyle {k_2\ov 2}$, equation \eqn{beta-gr} gives for
the running of the coupling $\l$ the same expression as the one in \eqn{resubeta}.

\section{Symmetry at the IR conformal point}

Our models provide an explicit example of an integrable smooth flow between exact CFTs. One of the end points is the sum of two WZW models at different levels.
In this section we investigate the other end point, that is the nature and the symmetries of the CFTs  to which the theory flows as one approaches the IR regime.
We will consider the two cases $(\l_1,\l_2)=(\l_0,\l_0)$ and $(\l_1,\l_2)=(\l_0,0)$ separately. In each case we will specify the corresponding CFT and its symmetries.

\subsection{CFT and its symmetries at $\l_1=\l_2=\l_0$}

 The corresponding CFT is obtained by setting  $\l_1=\l_2=\l_0 \d_{ab}$ in \eqref{defactigen}.
However, the symmetries of the CFT are more clearly exhibited if one uses the action before integrating out the fields $A_\pm$ and $B_\pm$.
In this case the action \eqn{gaufix} becomes
\be\label{CFT1}
\begin{split}
& S= S_{k_1}(g_1) +  S_{k_2}(g_2) +{k_1\ov \pi} \int d^2\s {\rm Tr}(A_- \del_+ g_1 g_1^{-1} -
B_+ g_1^{-1} \del_- g_1 + A_- g_1 B_+ g_1^{-1})
\\
&\quad +{k_2\ov \pi} \int d^2\s {\rm Tr}(B_- \del_+ g_2 g_2^{-1} -
A_+ g_2^{-1} \del_- g_2 + B_- g_2 A_+ g_2^{-1} - A_+ A_- - B_+ B_-)\ .
\end{split}
\ee
In order to clarify the nature of this CFT consider the following infinitesimal transformations
\be
\label{aaa}
\begin{split}
& \d A_\pm = -\del_\pm u_L + [A_\pm , u_L]\ ,\qq \d B_\pm = -\del_\pm u_R + [B_\pm , u_R]\ ,
\\
&\d g_1 = -u_L g_1 + g_1 u_R\  ,\qq \d g_2 = -u_R g_2 + g_2 u_L\ .
\end{split}
\ee
Then the variation of the action at the IR fixed point becomes
\be
\d S = {k_2-k_1\ov \pi} \int d^2\s {\rm Tr}(A_- \del_+ u_L + B_+ \del_- u_R)\ .
\ee
Hence, if $u_L=u_L(\s^-)$ and $u_R=u_R(\s^+)$, the action remains invariant.
We may investigate this in more detail by writing the would-be gauge fields $A_\pm$ and $B_\pm$ as
\be
A_\pm = \del_\pm h_\pm h_\pm^{-1}\ ,\qq B_\pm = \del_\pm k_\pm k_\pm^{-1} \ ,\qq h_\pm, k_\pm \in G\ .
\ee
The finite version of the transformation \eqn{aaa} for the $A_\pm$ and $B_\pm$ is equivalent to the following transformation for $h$ and $k$, namely
$h_\pm \to L^{-1}h_\pm$ and $k_\pm \to R^{-1} k_\pm$.
Then, with the aid of the Polyakov--Wiegmann formula we may write the action \eqref{CFT1} as
\be
\begin{split}
&S = S_{k_1}(h_-^{-1}g_1k_+) + S_{k_2-k_1}(k_-^{-1}g_2h_+)- S_{k_2}(h_-^{-1} h_+)
\\
& \qq + S_{k_1}(k_-^{-1}g_2h_+) - S_{k_2}(k_-^{-1}k_+)
\\
& \qq + S_{k_2-k_1}(h_-^{-1}) + S_{k_2-k_1}(k_+)\ .
\end{split}
\ee
The symmetry group of the CFT becomes transparent if we  change variables as $g_1\to g_1 h_+ k_+^{-1}$ and $g_2\to k_- h_-^{-1}g_2$. Then the action can be cast as
\be
\label{kkk}
\begin{split}
&S = S_{k_1}(h_-^{-1}g_1h_+) + S_{k_2-k_1}(h_-^{-1}g_2h_+)- S_{k_2}(h_-^{-1} h_+)
\\
& \qq + S_{k_1}(h_-^{-1}g_2h_+) - S_{k_2}(k_-^{-1}k_+)
\\
& \qq + S_{k_2-k_1}(h_-^{-1}) +  S_{k_2-k_1}(k_+)\ .
\end{split}
\ee
The first line is the gauged WZW action for the coset CFT
\be
\label{fhi3}
{G_{k_1}\times G_{k_2-k_1}\ov G_{k_2}}\Big |_L\ \otimes {G_{k_1}\times G_{k_2-k_1}\ov G_{k_2}}\Big |_R \ ,
\ee
indicating both the left and the right sectors.
The conformal invariance is generated by the transformations
\be
\begin{split}
&
h_+\to h_+ \Om(\s^-)\ ,\qq k_+\to k_+ \Om(\s^-)\ ,
\\
& h_-\to h_- \tilde \Om(\s^+)\ ,\qq k_-\to k_- \tilde \Om(\s^+)\ .
\end{split}
\ee
Under these the second and third lines of \eqn{kkk} generate two copies
of the current algebra for $G$ for the left and the right movers,  but at level zero. Hence,
for unitary representations this is trivial.

\no
The action \eqn{kkk} is also invariant under
\be
\begin{split}
& h_\pm \to L^{-1}(\s^-) h_\pm\ , \qq k_\pm \to R^{-1}(\s^+) k_\pm\ ,
\\
& g_1\to L^{-1}(\s^-) g_1 L(\s^-)\ ,\qq  g_2\to L^{-1}(\s^-) g_2 L(\s^-)\ .
\end{split}
 \ee
This generates the current algebra theory
\be
 G_{k_2-k_1}\big |_L\ \otimes  G_{k_2-k_1}\big |_R\  .
\ee
Combining the above and \eqn{fhi3} we obtain the following flow of CFTs
from the UV at  $\l_1=\l_2=0$  towards the IR at $\l_1=\l_2=\l_0$
\be
\label{sym-group}
G_{k_1}\times G_{k_2}
 \quad \overset{\text{IR}}{\Longrightarrow}
 \quad
  {G_{k_1}\times G_{k_2-k_1}\ov G_{k_2}} \times  G_{k_2-k_1}\ ,
\ee
one copy for the left and an identical one for the right movers.
This flow was speculated for the $SU(2)$ case in \cite{chiraliquids} based mainly on
symmetry arguments and further supported in \cite{Georgiou:2016zyo} for general groups based on the form of the anomalous dimensions of the current operators in the
CFT point in the IR.  Note that this flow is in accordance with Zamolodchikov's $c$-theorem \cite{Zamolodchikov:1986gt} since the central charge in the IR is smaller
than that in the UV.

\subsection{CFT and its symmetries at $\l_1=\l_0$, $\l_2=0$}

For the case of $\l_2=0$ the action  \eqref{defactigen} or equivalently \eqn{accctl2o} can be rewritten at the fixed point  $\l_{ab}=\l_0 \d_{ab}$,  by the use of the Polyakov-Wiegman identity as
\be
S=S_{k_1}(g_2 g_1) + S_{k_2-k_1}(g_2) \ .
\ee
This is the sum of two WZW actions with two copies of direct current algebra
$G_{k_1}\times G_{k_2-k_1}$ for the left and the right movers.
Hence, in this case, the theory smoothly flows from a CFT in the UV which is the sum of two WZW models, one with level $k_1$ and the other with level $k_2$ to another  CFT in the IR which is the sum of two WZW models, at levels  $k_1$ and $k_2-k_1$, respectively. That is
\be
G_{k_1}\times G_{k_2}
 \quad \overset{\text{IR}}{\Longrightarrow}
 \quad
  G_{k_1}\times   G_{k_2-k_1}\ ,
  \label{symm2}
\ee
one copy for the left and an identical one for the right movers.
The IR theory has indeed a smaller central change that the one in the UV, again in accordance with Zamolodchikov's $c$-theorem.

One might worry that our conclusion for the CFTs \eqn{sym-group} and \eqn{symm2} at the IR fixed
point could be an artifact of the large level approximation of our analysis.
However, the given answer
in terms of exact CFTs leaves no doubt that that the IR CFTs are the ones presented above but for finite level values.
The form of $\l_0$ in \eqn{levl0} for finite values of the levels may change, but not its very existence.

Our discussion was valid as long as $\l_0<1$. When $\l_0=1$, i.e. $k_1=k_2$, then the
IR fixed point seizes to exist and in fact the theory makes sense as long as we take an non-Abelian type
limit \cite{Georgiou:2017oly} of the PCM for $G\times G$.

\section{Discussions and future directions}

One of the intriguing features of two dimensional models is the existence of integrable quantum field theories interpolating between exact CFTs. The first example of such a flow was discovered in \cite{Zam-1,Cardy} and was realized via relevant perturbations of the unitary minimal models $\mathcal M_p$.  For one sign of the coupling constant the theory was argued to flow to another minimal model, namely $\mathcal M_{p-1}$. Subsequently, these flows were generalized, by applying thermodynamic Bethe ansatz techniques, to more general unitary minimal models involving coset spaces \cite{Zam-gen,Ravani} as well as to  integrable flows between non-unitary theories  of the type $\mathcal M_{p,q}$ \cite{Lassig:1991an,Ahn:1992qi} (see also \cite{Martins:1992ht,Ravanini:1994pt,Dorey:2000zb}). However, in all the aforementioned examples the description was based on integrability arguments and the theories were lacking a Lagrangian formulation.
In this paper, we explicitly constructed families of integrable $\s$-model actions smoothly interpolating between exact CFTs. Our realization uses two group elements of a general semi-simple group
and may have two or one distinct deformation matrices. 

Our construction resembles the similar construction of the doubly deformed integrable $\s$-models presented in \cite{Georgiou:2016urf} after allowing different levels for each of the WZW models.
Despite the fact that the methods of construction are similar, making the levels different has major
implications for the quantum behavior of the models. We have
computed the $\b$-function using gravitational methods and found that it exhibits a fixed point in the IR making our models particularly attractive.
Moreover, we proved the remarkable fact that the resulting theories are integrable in the case of isotropic couplings.  We explicitly constructed the Lax pairs and showed that the conserved charges are in involution.

Our models provide an explicit example of an integrable smooth flow between exact CFTs.
At the UV point the theory is described by the sum of two WZW models, one  at level $k_1$ and the other at level $k_2$. As soon as the perturbation is turned on our theories are driven towards another fixed point in the IR.
 When both coupling matrices are present,
 the IR  CFT is described by a coset CFT whose symmetry group is given by \eqref{sym-group}.
In the case of one coupling matrix the IR CFT is the sum of two WZW models one at level $k_1$ and the other at level $k_2-k_1$, \eqn{symm2}.
In both cases, the flow respects Zamolodchikov's $c$-theorem.

The motivation for the present paper was to find an action realizing all loop
effects  of the theory \eqn{pert2} including the existense of an IR fixed point.
This goal has been achieved since the $\b$-functions of our models do reproduce the all-loop $\b$-function of the deformation of the left-right asymmetric CFTs which was previously derived in \cite{LeClair:2001yp} and in \cite{Georgiou:2016zyo} using CFT methods and non-perturbative symmetries of the theory.
Although our realisation as a whole is left-right symmetric, as discussed below \eqref{expan1}, our models reproduce accurately not only the exact $\b$-function but also all the correlation functions of the left-right asymmetric model  of \eqref{pert2}.

A number of important questions still remains to be answered. It would be interesting to examine if one can construct a Lagrangian realization of the left-right asymmetric theories utilising a single group element and not two as we did in our construction. It would also be important to examine if the theories we constructed in this work can be embedded as solutions of type-II supergravity.  Since our theories are integrable this may result to
new integrable deformations of the $AdS_5 \times S^5$ superstring, for example. Along the same lines, it would be interesting to see if there are other choices for the deformation matrices, except the isotropic one, that preserve the integrability of the model.
Moreover, the fact that the $\l$ and $\eta$-deformations are related via Poisson-Lie T- duality and appropriate analytic continuations raises the question on the existence or not of new integrable models of the $\eta$-type which are dual to those constructed in the present work.
Finally, it would be interesting to analyze the implications of our Lagrangian description in the context of chiral liquids in one dimension \cite{chiraliquids} since such systems should be apparently described by left-right asymmetric theories.

\section*{Acknowledgments}

We acknowledge very useful discussions with K. Siampos.
G. Georgiou would like to thank the
Physics Department of the National and Kapodistrian U. of Athens
for hospitality during this project.


\end{document}